\documentclass[a4paper,10pt]{article}
\usepackage[utf8]{inputenc}
\usepackage{amsmath}
\usepackage{amssymb}
\title{  The Most General Form of   Deformation of the Heisenberg Algebra from the Generalized Uncertainty Principle}
\author{  Syed Masood$^1$, Mir Faizal$^{2, 3}$, Zaid Zaz$^4$,
\\ Ahmed Farag Ali$^{5,6}$, Jamil Raza$^1$, Mushtaq B Shah$^7$ \\  \\
$^1$Department of Physics,
International Islamic University,\\ H-10 Sector, Islamabad, Pakistan\\
$^2$Irving K. Barber School of Arts and Sciences, \\
University of British Columbia - Okanagan,\\
Kelowna, BC V1V 1V7, Canada \\
$^3$Department of Physics and Astronomy, University of Lethbridge, \\ Lethbridge, AB T1K 3M4, Canada \\
$^4$Department of Electronics and Communication Engineering , \\University of Kashmir,
Srinagar, Kashmir-190006, India
\\
$^5$Department of Physics, Faculty of Science, Benha University, \\Benha, 13518, Egypt
\\
$^6$Netherlands Institute for Advanced Study, \\ Korte Spinhuissteeg 3, 1012 CG Amsterdam, Netherlands.\\
$^7$ Department of Physics,  National Institute of Technology, \\
Srinagar, Kashmir-190006, India
}
\date{}
\begin{document}

\maketitle

\begin{abstract}
In this paper, we will propose the most general form of the deformation of Heisenberg algebra
motivated by the generalized uncertainty principle.
This deformation of the Heisenberg algebra will deform all quantum mechanical systems.
The form of the generalized uncertainty principle used to motivate these results will be motivated by the
space fractional quantum mechanics,
and non-locality in quantum mechanical systems.    We also analyse a specific limit of this generalized deformation
for one dimensional system,
and in that limit, a nonlocal deformation of the momentum operator generates a local deformation of all one dimensional
quantum mechanical systems.  We analyse the low energy  effects of this deformation on   a harmonic oscillator,
Landau levels, Lamb shift, and potential
barrier. We also demonstrate that this deformation leads to a discretization of space.
\end{abstract}

\section{Introduction}

A  universal prediction of almost all approaches to quantum gravity is  the  existence of a minimum measurable length scale,
and it is not possible to make physical measurements below this scale. String theory is one of the most important approaches to quantum gravity.
 The string length scale acts as a minimum length scale in string theory as the strings are the smallest probes that exist in the
 perturbative string theory \cite{z2}-\cite{2z}.  The existence of a minimum measurable length in loop quantum gravity turns the
 big bang into a big bounce \cite{z1}. It can be argued from black hole physics that any theory of quantum gravity should have a minimum
 measurable length scale of the order of the Planck scale   \cite{z4}-\cite{z5}. This is because the energy needed to probe any region of space below
 Planck scale is larger than the energy required to form a mini black hole in that region of space.
Even though the existence of a minimum measurable length scale is predicted from various different theories, the existence of a minimum measurable length
scale is not consistent with the  usual   Heisenberg uncertainty principle. This is because according to the usual Heisenberg uncertainty principle,
length can be measured with arbitrary precision, as long as the  momentum is not measured. To incorporate  the existence of a
minimum measurable length scale in the uncertainty principle, the usual   Heisenberg uncertainty principle
has to be generalized to a generalized uncertainty
principle  (GUP)  \cite{1}-\cite{15}. The  uncertainty principle is related to
the Heisenberg algebra, and so any  modification of the uncertainty principle will deform the
Heisenberg algebra \cite{17}-\cite{53}. The deformation of the Heisenberg algebra will in turn modify the
  the coordinate
representation of the momentum operator \cite{18}-\cite{10}.
As the coordinate representation of the momentum operator is used to derive the
quantum mechanical behavior of a system, the modification of the coordinate
representation of the momentum operator will   produce correction terms for all quantum mechanical systems.
It may be noted that even though the minimum measurable length scale has to exist at least at the Planck scale, it is
possible for the minimum measurable length scale to exist at a much lower length scale. In fact, it has been demonstrated that if the minimum measurable
length  scale exists at a scale much lower than the Planck scale, then the deformation of the Heisenberg algebra produced by it can have interesting
low energy consequences \cite{54}.

Even though the  generalized uncertainty principle is motivated by the existence of a minimum measurable length scale,
there can be other motivations
for studying the theories based on the  generalized uncertainty principle. It has been demonstrated that the
generalized uncertainty principle can be motivated
from  the breaking of  supersymmetry in supersymmetric field theories . It  is important to break supersymmetry at
sufficient large energy
scale because the low energy supersymmetry has not been observed. Even though there are
various different mechanisms for breaking
supersymmetry, it has been demonstrated that the breaking of supersymmetry due to non-anticommutativity   deforms the
Heisenberg algebra, and this deformed Heisenberg algebra is consistent with the existence of the
generalized uncertainty principle \cite{field}.
The coordinate representation of the momentum operator produced from this deformation of the Heisenberg algebra, and the
coordinate representation of the momentum operator produced from minimum measurable length scale contains a quadratic power of momentum (at the leading order).
However, it is also possible to motivate a different  deformation of the Heisenberg algebra, and this deformation of the Heisenberg algebra occurs
is due to the   doubly special relativity
 \cite{2}-\cite{3}. The doubly special relativity is a theory in which the Planck energy and the
  velocity of light are universal constants, and as the theory contains more than one universal constant, is called
  the doubly special relativity. The doubly special relativity is motivated from the deformed
  energy-momentum dispersion relation which occurs due to the existence of a maximum energy scale. Such
 a deformation of the energy-momentum dispersion relation occurs in various different approaches to quantum gravity,
 such as the  discrete spacetime \cite{1q},
the spontaneous symmetry breaking of
Lorentz invariance in string field theory \cite{2q}, spacetime foam models \cite{3q}, spin-network in loop quantum gravity \cite{4q},
non-commutative geometry \cite{5q}, and Horava-Lifshitz gravity \cite{6q}. It is possible to combine the
quadratic deformation of the
  Heisenberg algebra  (motivated by the  existence of a minimum measurable length and breaking of supersymmetry),
  with the deformation
  of the Heisenberg algebra produced by the  doubly special relativity  \cite{n4}-\cite{n5}.
  The coordinate representation of the momentum operator for such a deformed Heisenberg algebra which is produced by the combination
  of both these deformations contains linear powers of the momentum operator in the coordinate representation of the momentum
  operator. This produces fractional derivative  contributions in any dimension beyond the simple
  one dimensional case. However, it is possible to study these  fractional derivative terms using
  the harmonic extension of functions  \cite{mir}-\cite{m1}.

  One of the most interesting consequences of the deformed Heisenberg algebra (containing linear powers of momentum in the coordinate
  representation of the momentum operator) is that it leads to a discretization of space \cite{n4}.
  It may be noted that it is possible to have low energy consequences of this  deformation of the  Heisenberg algebra,
  if the deformation scale is assumed to be sufficient large \cite{54}.  In fact, it has been demonstrated that for simple quantum mechanical systems
  like the harmonic oscillator, the
Lamb shift and the Landau levels get corrected by this deformed Heisenberg algebra, and these corrections can be experimentally measured
\cite{n7}.  It may be noted that second quantization of deformed fields theory  has been studied, and the deformed field theories have
been motivated by the  generalized uncertainty principle   \cite{n9}-\cite{mir1}. As interesting physical consequences have been obtained using
the deformation of the momentum operator by both the linear and quadratic form of the generalized uncertainty principle,
we will propose the most general form of such a deformation,  and we will analyse an interesting limit of this most generalized uncertainty
principle.

\section{Generalized Uncertainty Principle}
In this section, we will propose the most general form of the deformation of the momentum operator,
and the effect it can have on different quantum mechanical
systems.  The modification of the  usual uncertainty principle to a generalized uncertainty principle is motivated
from the existence of minimum measurable length scale \cite{1}-\cite{15}, double special relaivity \cite{2}-\cite{3},
spontantious symmetry breaking  \cite{field},
string theory \cite{z2}-\cite{2z}, loop quantum gravity  \cite{z1}, black hole physics \cite{z4}-\cite{z5},
 and modified dispersion relation which occurs in discrete spacetime \cite{1q},
the spontaneous symmetry breaking of
Lorentz invariance in string field theory \cite{2q}, spacetime foam models
\cite{3q}, spin-network in loop quantum gravity \cite{4q},
non-commutative geometry \cite{5q},  Horava-Lifshitz gravity \cite{6q}.
In the simple case of a one dimensional
generalized uncertainty principle,
the usual uncertainty between momentum $\Delta p$ and position $\Delta x$ is modified from its usual form
$\Delta p \Delta x \geq \hbar/2$ to a deformation by some    function of $p$, for example,
$\Delta p \Delta x \geq \hbar/2+  \hbar \lambda (\Delta p )^2 $,
where $\lambda$ is the deformation parameter. Such a
deformation has been considered for higher dimensions \cite{mir}.
However, uncertainty principle is closely related to the Heisenberg algebra, so a deformation of the
uncertainty principle will deform the Heisenberg algebra. However, almost all the work done
on the deformed Heisenberg algebra has been done on the deformation motivated from generalized
uncertainty principle containing
a quadratic momentum term   \cite{1}-\cite{15} and a linear momentum term  \cite{n4}-\cite{n5}.
In this paper, we will first construct the most general deformation of the Heisenberg algebra,
and then analyse a specific limit of this algebra. Even though a lot of work has been done on both
linear and quadratic
deformation of the Heisenberg algebra, such a limit of this deformation has not been analysed.
Now we can also write the most
general deformation of the  Heisenberg algebra  as
\begin{eqnarray}
 [x^i, p_j] = i \hbar \left[\delta^i_j + f[p]^i_j\right],
\end{eqnarray}
where $f[p]^i_j$ is a suitable tensorial function    that is  fixed by the form of the generalized uncertainty
principle, and which in turn fixes the form of coordinate representation of
 the momentum operator.
The deformation of the Heisenberg algebra in turn deforms the coordinate representation of the momentum operator.
It may be noted that for  for a quadratic generalized uncertainty principle,
 the coordinate representation of the momentum operator gets deformed from
 $p_i = - i\hbar \partial_i  $ to $\tilde p_i = - i\hbar \partial_i  (1 - \lambda \hbar^2 \partial^j \partial_j)$,
where $\lambda$ is the deformation parameter \cite{18}-\cite{10}.
Thus, as the  original moment
momentum is   $p_i = - i\hbar \partial_i $, the quadratic generalized uncertainty principle deforms the momentum to
\begin{eqnarray}
 p_i \to \tilde p_i = p_i (1+ \lambda p^j p_j)
\end{eqnarray}
We can define this deformation for a one dimensional case as follows,
$ p \to \tilde p =  p (1+ \lambda p^2)
$. Now we can write the deformation of a one dimensional  quantum mechanical  Hamiltonian
for a particle   as
\begin{eqnarray}
 H = \frac{p^2}{2m} + V(x) \to  H + \lambda H_{h},
\end{eqnarray}
where the correction term  scales as  $ H_{h} \sim p^4 $.

The deformation   produced by combining this quadratic deformation with doubly special relativity deforms the
coordinate representation of the momentum operator from $p_i = - i \hbar \partial_i$ to
$\tilde p_i = -i  (1 -  \lambda_1 \hbar\sqrt{-   \partial  ^j \partial_j} - 2 \lambda _2    \hbar^2 \partial ^j
  \partial _j ) \hbar \partial _i$
\cite{n4}-\cite{mir}. Thus, the effect of this deformation is that the original momentum $p_i = - i \hbar \partial_i$, gets deformed to
\begin{eqnarray}
p_i  \to \tilde p_i = p_i \left(1 +  \lambda_1  \sqrt{p^j p_j} + 2 \lambda _2  p^j p_j\right).
\end{eqnarray}
It may be noted that in the deformation produced by the combination  of the quadratic deformation with doubly special relativity $\lambda_2 = 2 \lambda_1^2$
\cite{n4}-\cite{mir}.
Now we can write this deformation for a one dimensional system
\begin{eqnarray}
 p \to \tilde p = p (1+ \lambda_1 p + \lambda_2 p^2).
\end{eqnarray}
The deformation  of a quantum mechanical    Hamiltonian for a particle in one dimension by this form of generalized uncertainty can now be written as
\begin{eqnarray}
 H = \frac{p^2}{2m} + V(x) \to  H + \lambda_1 H_{h1} + \lambda H_{h2}
\end{eqnarray}
where the correction terms scale as
 $ H_{h1} \sim p^3 $ and $ H_{h2} \sim p^4$.

 It may be noted that in higher dimensions linear contributions from momentum operator introduce   fractional derivative terms.
It may be noted that in any dimension greater than the simple one dimensional case,
such   fractional derivative terms will occur for any  power of momentum  in the deformation of the momentum operator.
This is because for any   power of the momentum operator $ p_i \to \tilde p_i = p_i (1 + \lambda_r (p^j p_j)^{(r/2)} $, we can
write the coordinate representation as $\tilde p_i = - i\hbar \partial_i (1 +\lambda_r (-\hbar^2 \partial^i   \partial_i)^{(r/2)} )$.
Now when  $r = 2n$, then this term does not contain   fractional derivative terms $(\hbar^2 \partial^i  \partial_i)^{r/2}
= (\hbar^2 \partial^i   \partial_i)^n$. However, when $r = 2n + 1 $, then this term   contains   fractional derivative terms $(\hbar^2 \partial^i
\partial_i)^{r/2}
= (\hbar^2 \partial^i  \partial_i)^n (\hbar^2  \partial^i  \partial_i)^{1/2}$. Such fractional derivative terms can be effectively analysed using
harmonic extension of functions \cite{mir}.
However, it is also possible to analyse any fractional derivative term using the theory of harmonic extension of functions,
and so we can also propose that this deformation contains arbitrary fractional powers of the momentum and write
\begin{eqnarray}
 p_i  \to \tilde p_i =  p_i \left(1 +  \sum  \lambda_{1r} (p^j p_j)^{r/2} \right)
\end{eqnarray}
It may be noted that such fractional derivative terms occur in space fractional quantum mechanics \cite{frac}-\cite{frac2}.
In this equation  the Brownian trajectories in Feynman path
integrals are replaced by Levy flights. It is possible to study Levy crystals in condensed matter physics using such
a fractional quantum mechanics \cite{quant}. Fractional quantum mechanics has also been applied in optics, and
this is because fractional quantum harmonic oscillator have been used to analyse
  dual Airy beams which can  be selectively generated under off-axis longitudinal pumping \cite{optics}.
  Thus, there is a good motivation to incorporate such terms in the generalized uncertainty principle.
So, we can also include $(p^i p_i)^{r}$ terms in the generalized uncertainty principle.

It may be noted that this deformation of the momentum operator can produce   fractional derivative terms.
Let us consider a simple deformation of the momentum operator involving fractional derivative terms,
$ p_i \to \tilde p_i = p_i  (1 +     \lambda (p^j p_j)^{1/2} )$, and   in this deformation  the Schroedinger's equation will contain a
  fractional derivative term of the form $\sqrt{-\partial^j\partial_j}$.
Even though such fractional derivative   terms  exist in the Schroedinger's equation, it is possible to deal with them using harmonic extension of function.
Thus, we will formally analyze   $i ({\partial^i \partial_i}  )^{1/2}$   using
 the harmonic
extension of wave function from  $\mathbb{R}^3 $ to  $\mathbb{R}^3 \times
\left( 0, \infty \right)$ \cite{hamr}-\cite{hamr1}. So, let  $u:\mathbb{R}^3 \times
\left( 0, \infty \right)\mathbb{\longrightarrow }\mathbb{R}$ be a     harmonic function which is the harmonic extension of
a wave function $\psi :\mathbb{R}^3 \mathbb{%
\longrightarrow }\mathbb{R}$, such that the  restriction of $u$ to $\mathbb{R}^3 $ coincides with  $\psi$.
Now  this can be analyzed as a
Dirichlet problem, which is given by
\begin{eqnarray}
u\left( x,0\right) =\psi\left( x\right), &&
\nabla _4^{2}u\left( x,y\right) =0.
\end{eqnarray}%
Here  $\nabla _4^{2}$ is the Laplacian operator in $\mathbb{R}^{4}$, such that  $%
x\in \mathbb{R}^3 $ and $y\in \mathbb{R}$.  It may be noted  that there  is a unique harmonic extension  $u\in C^ \infty (\mathbb{R}^3 \times
\left( 0, \infty \right))$,   for any smooth function on $C_0^ \infty (\mathbb{R}^3 )$. So, we
  can analyze the action of the differential operator $ i(\partial^i \partial_i)^{1/2}$   on the wave functions $\psi:\mathbb{R}^3 %
\mathbb{\longrightarrow }\mathbb{R}$ using the harmonic extension of functions. Now  as   $u:$ $%
\mathbb{R}^3 \times
\left( 0, \infty \right)\mathbb{\longrightarrow }\mathbb{R}$,
is the     harmonic extension of the wave function, we can write
\begin{eqnarray}
\left(  i(\partial^i \partial_i)^{1/2}\psi\right) \left( x\right) =-\left. \frac{\partial u\left( x,y\right)
}{\partial y}\right\vert _{y=0}.
\end{eqnarray}%
The function   $\left(  i(\partial^i \partial_i)^{1/2}\psi\right) \left( x\right) $ also has   harmonic
extension to $\mathbb{R}^3 \times
\left( 0, \infty \right)$. This harmonic extension  can be  written as  $u_{y}\left(
x,y\right) $, when   $u\left( x,y\right) $ is the harmonic extension of $\psi(x)$.
 So,  from the successive applications of $ i(\partial^i \partial_i)^{1/2}$, we obtain
\begin{eqnarray*}
\left(  i(\partial^i \partial_i)^{1/2}\left(  i(\partial^i \partial_i)^{1/2}\psi\right) \right) \left( x\right) &=&\left. \frac{\partial
^{2}u\left( x,y\right) }{\partial y^{2}}\right\vert _{y=0} \nonumber\\
&=&
\left. -\nabla _3 ^{2} u\left( x,y\right) \right\vert _{y=0}\nonumber \\ &=&-\nabla _3 ^{2}\psi\left(
x\right).
\end{eqnarray*}%
Thus, we can write   $
[ i(\partial^i \partial_i)^{1/2}]^{2}\psi ( x ) = ( -\nabla _3 ^{2} ) \psi ( x )
$, and give a formal  meaning to the fractional differential operator as
$ i(\partial^i \partial_i)^{1/2}= ( -\nabla _3 ^{2} ) ^{1/2}
$.  It may be noted  that  for    $u\in C^{2}\left( \mathbb{R}\times
\left( 0, \infty \right))\right) $, we can write
\begin{eqnarray*}
\left(  i(\partial^i \partial_i)^{1/2}\left( \partial _{i}\psi\right) \right) \left( x\right)  &=&-\left.
\partial _{y}\left( \partial _{i}u\left( x,y\right) \right) \right\vert
_{y=0} \\
&=&-\partial _{i}\left. u_{y}\left( x,y\right) \right\vert _{y=0} \\
&=&\partial _{i}\left(  i(\partial^i \partial_i)^{1/2}\psi\right) \left( x\right).
\end{eqnarray*}%
So,  we obtain
\begin{eqnarray}
\left( -\nabla _3 ^{2}\right) ^{1/2}\partial _{i}=\partial
_{i}\left( -\nabla _3 ^{2}\right) ^{1/2}.
\end{eqnarray}
Thus, this fractional derivative commutes with the usual derivatives.

It is possible to demonstrate the this fractional derivative operator, $ i(\partial^i \partial_i)^{1/2}$,  is an  self-adjointness   operator
 \cite{ha}-\cite{5a01}.  Now let   $
u_1\left( x,y\right) $ and $u_2\left( x,y\right) $ be the harmonic extensions of  $\bar \psi_1\left( x\right) $ and
$\psi_2\left( x\right) $, respectively.  Furthermore, let    both of these   harmonic
extensions vanish for $\left\vert x\right\vert ,\left\vert y\right\vert
\longrightarrow \infty $. Now  we can write  \cite{5a01}
\begin{eqnarray}
&&\int_{\mathcal{C}} \nabla _4u_1\left( x,y\right)  \cdot  \! {\partial^i \partial_i}
_{4}u_2\left( x,y\right) dxdy \!\!\! \nonumber \\ &=&
\int_{\mathcal{C}}\!\!\nabla _4\! \cdot\!  \left( u_1\left( x,y\right) {\partial^i \partial_i}
_{4}u_2\left( x,y\right) \right) dxdy  \notag \\
&=&\int_{\partial \mathcal{C}}u_1\left( x,y\right) \nabla _4u_2\left(
x,y\right) ~dxdy  \notag \\
&=&-\int_{\mathbb{R}^3 }\left. u_1\left( x,y\right) \frac{\partial }{\partial
y}u_2\left( x,y\right) \right\vert _{y=0} dx,   \label{integral over C}
\end{eqnarray}%
where  $\partial \mathcal{C}$  is the border of $\mathcal{C}$.
So,  for  harmonic extensions $u_1 $ and $ u_2$, we can write  \cite{ar}
\begin{eqnarray*}
\int_{\mathcal{C}}u_1\left( x,y\right) \nabla _4^{2}u_2\left( x,y\right)
~dxdy-\int_{\mathcal{C}}u_2\left( x,y\right) \nabla _4^{2}u_1\left(
x,y\right) ~dxdy=0.
\end{eqnarray*}%
Now this can be written as
\begin{eqnarray*}
\int_{\mathbb{R}^3 }\left. \left( u_1\left( x,y\right) \frac{\partial }{%
\partial y}u_2\left( x,y\right) -u_2\left( x,y\right) \frac{\partial }{\partial y%
}u_1\left( x,y\right) \right) \right\vert _{y=0}~dx=0.
\end{eqnarray*}%
We can write this in terms of  $\bar \psi_1\left( x\right) $ and $\psi_2 \left( x\right) $  as
\begin{eqnarray*}
\int_{\mathbb{R}^3 }\left( \bar \psi_1\left( x\right) \frac{\partial \psi_2\left( x\right)
}{\partial y}-\frac{\partial \bar \psi_1\left( x\right) }{\partial y}\psi_2\left( x\right)
\right) ~dx=0.
\end{eqnarray*}%
So, we obtain
\begin{eqnarray}
\int_{\mathbb{R}^3 }\bar \psi_1\left( x\right)  i(\partial^i \partial_i)^{1/2}\psi_2\left( x\right) dx=\int_{%
\mathbb{R}^3 }\psi_2\left( x\right)  i(\partial^i \partial_i)^{1/2}\bar \psi_1\left( x\right) dx.
\label{self adjointness}
\end{eqnarray}
Thus, we can deal with the fractional derivative terms produced by the deformation of the momentum operator by the    generalized uncertainty principle using
harmonic extension of wave function.
It may be noted that it is known that such fractional derivative terms are self-adjointness   operator \cite{hamr}-\cite{5a01}. However, in this paper, we
have proposed them to be produced by a deformation of the generalized uncertainty principle.  It may be noted that the  self-adjointness of the
momentum operator deformed by generalized uncertainty principle has been analysed over different  different domains \cite{diffe}. In this paper,
we will first propose the most general form of such a deformation, and then analyze a specific interesting deformation produced by the generalized uncertainty
principle.

\section{Non-Locality}
It is also possible to analyse a more general deformation of the momentum operator, which would contain inverse powers of the momentum
operator. Such a deformation can be motivated from non-local quantum mechanics.
Now for example the Schroedinger equation with a non-local term can be written as \cite{linear}-\cite{linear6}
\begin{equation}
 i \hbar \partial_t \psi(x)  + \frac{1}{2m} \hbar^2 \partial^i \partial_i   \psi (x)  - V(x) \psi (x) =     \int d^3 x' K(x, x')\psi (x'x)
\end{equation}
where   $K(x, x')   $ is a non-local operator,  and such nonlocal terms are written as functions of $(p^i p_i)^{-1}$. In fact,
this can be easily seen for a very simple  non-local deformation of a scalar field theory. Nonlocal deformation of field theory
has been studied using  axiomatic field theory \cite{fieldtheory}-\cite{fieldtheory2}. Non-local deformation of gravity has also been studied,
and  such models of non-local gravity have been used to produce interesting physical results \cite{gravity}-\cite{gravity1}. Nonlocal deformation
of scalar field theory has also been studied \cite{scalar1}-\cite{scalar2}.
However, we will only consider a very simple nonlocal deformation of  a simple massless scalar field theory, whose equation of motion
\begin{eqnarray}
   \hbar^2  \partial^\mu  \partial_\mu \psi (x) = 0,
\end{eqnarray}
will be deformed by  a non-local source term,
\begin{eqnarray}
 \hbar^2 \partial^\mu  \partial_\mu \psi (x) =      \lambda \int d^4 y G(x-y) \psi (y),
\end{eqnarray}
where $\lambda$ is the coupling parameter which measures the coupling of the nonlocal part of this theory, and
\begin{eqnarray}
 G(x-y) =  \int  \frac{d^4 p}{(2\pi)^4}    \frac{1}{p^2} \exp i p . (x-y).
\end{eqnarray}
where $p^2= p^\mu p_\mu$, and its spatial part is $p^i p_i$.
 Now this can be written as
 \begin{eqnarray}
 &&  - \int d^4y [ \delta(x-y) \hbar^2 \partial^\mu  \partial_\mu  \psi (y) - \lambda G(x-y) \psi (y)]  \nonumber \\&=&
 \int \frac{d^4 p d^4y}{(2\pi)^4}  \left[      \left(p^2 +  \frac{\lambda}{p^2} \right) \exp i p . (x-y) \right]\psi (y) \nonumber \\ &=&
      0.
\end{eqnarray}
Hence, this non-local deformation of scalar field theory will be produced by the following deformation of the four momentum
\begin{eqnarray}
 p^2 \to \tilde p^2 =  p^2 + \frac{\lambda}{p^2}.
\end{eqnarray}
If we consider the temporal deformation, we get an extra term  of the form $p^{-2} = (p^\mu p_\mu)^{-1}$. However,
by neglecting the temporal deformation, we are only left with spatial deformation of the form,
$(p^i p_i)^{-1}$.
Thus, we can write the most general form
of generalized uncertainty principle by taking such inverse powers into account,
\begin{eqnarray}
 p_i \to \tilde p_i = p_i \left(1 +  \sum  \lambda_{1r} (p^j p_j)^{r/2}    +  \sum  \lambda_{2r} (p^j p_j)^{-r/2}  \right)
\end{eqnarray}
here $\lambda_{1i}$ and $\lambda_{2i}$ are suitable coefficients. It may be noted that it is possible to consider both positive and negative
values of $\lambda_{1i}$ and $\lambda_{2i}$. In fact, both positive and
negative values of such coefficients for generalized uncertainty principle have been considered
in analyzing the effects of generalized uncertainty principle on the thermodynamics of black holes
\cite{thermo}-\cite{thermo1}. However, if we want to  impose the condition that $\lambda_{1i}>0$ and $\lambda_{2i}>0$,
then we can write the most general form of deformation of the momentum operator as
\begin{eqnarray}
 p_i \to \tilde p_i =  p_i \left(1 \pm \sum  \lambda_{1r} (p^j p_j)^{r/2}    \pm  \sum \lambda_{2r} (p^j p_j)^{-r/2}  \right).
\end{eqnarray}
This is the most general form of deformation of the momentum operator that can be constructed, and it would be interesting to analyse specific
limits of this deformation. Here  we have included both even and odd powers of momentum in this deformation. However, the Hamiltonians with
odd powers of momentum will   violate  parity, and this can also have interesting physical consiquences.

It may be noted that  there is a interesting
 non-local deformation of the momentum operator, such that the one-dimensional Hamiltonian remains local.
Now let us consider this  simple limit of this general deformation of the momentum operator.
So, this  limit will contain a non-local term in the deformed momentum operator, but the Hamiltonian
for a one dimensional particle constructed from such a deformed momentum operator will not contain any
such non-local term. This can be achieved if we consider the following the deformation of the  momentum operator
for a one dimensional system,
\begin{eqnarray}
 p \to \tilde p = p \left(1 + \frac{\lambda }{ p}\right).
\end{eqnarray}
This will deform the usual Hamiltonian as
\begin{eqnarray}
 H = \frac{p^2}{2m} + V(x) \to  H + \lambda H_{h},
\end{eqnarray}
where
\begin{eqnarray}
\lambda H_{h} = \frac{\lambda p}{m}.
\end{eqnarray}
So, the new deformation scales as $H_h \sim p $, and such a linear term in Hamiltonian is a totally new deformation.
We will now analyse its effect on simple quantum mechanical system. Now for this deformation by a  $p^{-1}$ term,
    the Hamiltonian is give by a sum of   self-adjoint operators, and so this deformation  is well defined.
    Furthermore, this deformation   produces a odd power of momentum in the Hamiltonian, and so this Hamiltonian violates parity.
It may be noted that even though the quadratic and linear generalized uncertainty principle has
been motivated from minimum measurable length, and this length exists at Planck scale due to quantum gravitational effects,
it is possible to take the minimum measurable length scale at an intermediate length scale between Planck scale and electroweak scale,
and such a consideration can have  low energy consequences \cite{54}. In this paper, we will analyse the generalized uncertainty principle
using non-local deformation of the momentum operator,
and so we cannot directly relate this form of generalized uncertainty principle to the scale at which minimum length exists.
However, we can still use the available experimental data to fix a bound on $\lambda$.
Thus, if such a  nonlocal deformation of the coordinate representation  of the momentum operator  exists at a scale
beyond the available experimental data, then such an effect can be used to detect using the results obtained in this paper.

\section{Length Quantization}
One of the most interesting results of the deformation of momentum operator by a linear term is that it
predicts the discretization
of space. It may be noted that it has been demonstrated that such a result
occur for a deformation of the Heisenberg algebra motivated by the generalized uncertainty principle containing
a linear term in momentum \cite{n4}. This is because the box can only contain a particle, if the box is a multiple of
some fundamental length scale. This fundamental length scale does not depend on the length of a box, and as this holds
for a  box of an arbitrary length, it was proposed that all length in nature will be a multiple of this fundamental length
scale. Thus, this deformation produced a discrete structure for space.
In fact, the generalization of such a result to a relativistic Dirac equation has also
been done, and it was observed that even in this case the space gets a discrete structure \cite{main2}.
However, such a effect does not occur for the deformation motivated by the the generalized uncertainty principle containing
a quadratic term in momentum.
We will demonstrate that such an effect also occur due to the deformation of the momentum operator
by $p \to p (1 + \lambda p^{-1})$. The  deformation of a Schroedinger equation for a free particle, can be written as
 \begin{eqnarray}
 	\frac{d^{2} \psi}{dx^{2}} +\Big(\frac{2\iota\lambda}{\hbar}\Big)\frac{d\psi}{dx} + \frac{2mE}{\hbar^{2}}=0
 \end{eqnarray}
The solution to this deformed Schroedinger equation is given  by
\begin{eqnarray}
\psi &=& Ae^{\frac{\iota}{\hbar}\Big[\sqrt{\lambda^{2}+2mE} -\lambda\Big]x} + Be^{\frac{-\iota}{\hbar}\Big[\sqrt{\lambda^{2}+2mE} + \lambda\Big]x}
\nonumber \\
 &=& Ae^{\frac{\iota k_{1} x}{\hbar}} + Be^{\frac{-\iota k_{2} x}{\hbar}}
\end{eqnarray}
where $k_{1}=\sqrt{\lambda^{2}+ 2mE} - \lambda$  and $k_{2} = \sqrt{\lambda^{2}+ 2mE} + \lambda$.
Now the following boundary conditions hold for a  particle in a box,
$x=0$, $\psi=0$  and at $x=L$,  $\psi=0$. Thus, applying the first boundary condition,
$x=0$, $\psi=0$  we get $A = -B$, so we can write
\begin{eqnarray}
\psi= A\Big(e^{\frac{\iota k_{1} x}{\hbar}} - e^{-\frac{\iota k_{2}  x}{\hbar}}\Big).
 \end{eqnarray}
Applying the second boundary condition  $x=L$, $\psi=0$, we obtain
 \begin{eqnarray}
 A\Big(e^{\frac{\iota k_{1} L}{\hbar}} - e^{-\frac{\iota k_{2}L  }{\hbar}}\Big)=0.
  \end{eqnarray}
Now as $A\neq 0$, we can write
 \begin{eqnarray}
 e^{\frac{\iota(k_{1}+k_{2})L}{\hbar}}=1.
 \end{eqnarray}
Thus, we obtain
 \begin{eqnarray}
 \frac{(k_{1}+k_{2})L}{\hbar}=2n\pi.
 \end{eqnarray}
 So, the length of the box can be expressed as
 \begin{eqnarray}
 L=\frac{n2\pi\hbar}{k_{1}+k_{2}}
 \end{eqnarray}
 Now using the    values of $k_{1}$ and $k_{2}$, we obtain
 \begin{eqnarray}
 L=\frac{n\pi\hbar}{\sqrt{\lambda^{2}+2mE}}.
 \end{eqnarray}

 Thus, no   particle can exist in the box,  if the length of the box is not  quantized in terms of this discrete unit.
 However, as the box is of arbitrary length, this suggests that all lengths in space are quantized in terms of this
 discrete unit.
 Thus, this deformation of the momentum operator predicts the discretization of space.
 It may be noted that a similar result about length quantization was obtained using the generalized uncertainty principle
 with a linear momentum term  \cite{n4}-\cite{main2}. So, what we have demonstrated is that the effect of the $p^{-1}$ deformation
 is the quantization of length, and a similar effect can also be generated from a the generalized uncertainty principle
 with a linear momentum term  \cite{n4}-\cite{main2}.
 It may also be noted   that for the deformation by a linear term,
 the unite of this discretization did not depend on the energy of the probe.
 However, for the deformation produced by $p^{-1}$ term, the unite of discretization depends on
 the energy of the particle used to probe it.
Thus, we obtain a geometry, where the structure of space  depends on the energy of the probe. It may be noted that the
  gravity's rainbow has been constructed by assuming that the geometry of spacetime
  depends on the energy of the probe \cite{rainbow0}-\cite{rg}.
The gravity's rainbow can  be motivated from the string theory \cite{gr}. This is because the
  constants in a field theory flow due to the
  renormalization group flow, and so they depend on the scale at which a field theory would be probed. However, the scale at which a theory will be
  probed would depend on the energy of the probe. Thus, as the constants in a field theory   depend explicitly on the scale at which such a theory is  probed,
  they   also depend implicitly on the energy of the probe. Now it is also known that the
  string theory can be regarded  as a two dimensional conformal field theory,  and  the  target
space metric can be regarded  as a matrix of coupling constants of this two dimensional conformal field theory.
Thus, the target space  metric will also flow due to the renormalization group flow. This  would make the metric of the spacetime
depend on the energy of the probe producing gravity's rainbow.   Now as the
string theory has also been used as a motivation for the generalized uncertainty principle  \cite{z2}-\cite{2z},
it was expected that  a
certain forms of generalized uncertainty principle could produce   similar results.

Here we have been able to demonstrate that this particular form of generalized
uncertainty principle   makes  the microscopic structure of space depend on the energy of the probe.
So, it is possible that such a deformation can change the macroscopic structure of spacetime,
 and make it energy dependent. However, to construct such a theory, we would first have to analyse such an effect on curved spacetime.
 It has been demonstrated that the  deformation by a linear momentum term also leads to a discreteness of  space,  even when a weak gravitational field
 is present \cite{gravv}. It would be interesting to carry out such calculations for the deformation  by a $p^{-1}$ term.
It is expected that again the unite of discretization will depend on the energy of the probe. Then it might be possible to analyse the first
order corrections to the macroscopic geometry, due to this energy dependent discreteness. It would then be
possible to absorb such energy dependence in the metric, and
this would make the metric energy dependent, and we will be able to obtain results similar to  gravity's rainbow.   It may also be noted
that it has been demonstrated the generalized uncertainty principle in curved spacetime lead to a deformation  of the equivalence principle \cite{ep}-\cite{ep12}, 
and doubly special relativity (which is the main motivation for gravity's rainbow) is also based on the modification of the equivalence
principle  \cite{2}-\cite{3}. This is another reason to expect that a certain form of generalized uncertainty principle could produce results similar to
the gravity's rainbow.

\section{Harmonic Oscillator}
In this section, we will analyse the effect of this deformation on a harmonic oscillator.
The harmonic oscillator is important as it forms a toy model for various different physical systems.
The Hamiltonian for the harmonic oscillator gets deformed by this generalized uncertainty principle as
The deformed Hamiltonian for harmonic oscillator is
\begin{eqnarray}
	H &=&  \frac{p^{2}}{2m} + \frac{k x^{2}}{2}  \nonumber \to
	\frac{p^{2}}{2m} + \frac{k x^{2}}{2} + \frac{\lambda p}{m}.
\end{eqnarray}
The  first order correction to the  ground state of this harmonic oscillator is given by
\begin{eqnarray}
\Delta E_{0} &=&\int_{-\infty}^{+\infty}{\psi_{0}}^{*} \Big(\frac{\lambda p}{m}\Big) \psi_{0} dx\nonumber \\
&=& \frac{-\iota \hbar\lambda}{m}\int_{-\infty}^{+\infty}\psi_{0} \frac{d}{dx} (\psi_{0})dx,
 \end{eqnarray}
where $\psi_0$ is the ground state wave function of the original harmonic oscillator (without any contribution from $\lambda P/m$),
and it is given by (with $\alpha = m\omega/ 2 \hbar$),
\begin{equation}
 \psi_{0}=\Big(\frac{m\omega}{\pi\hbar}\Big)^\frac{1}{4} e^{-\alpha x^2}.
 \end{equation}
Now using $d\psi_0/dx  =  -2 \alpha x \psi_0  $, we obtain
 \begin{eqnarray}
 \Delta{E_{0}}&=&\Big(\frac{m\omega}{\pi\hbar}\Big)^\frac{1}{2}\frac{\lambda(-\iota\hbar)}{m}\int_{-\infty}^{+\infty}e^{-2\alpha x^2}(-2\alpha x) dx
 \nonumber \\ &=& 0.
 \end{eqnarray}
 Thus, there   is no effect of this deformation on the  ground state energy of a harmonic oscillator at first order of the perturbative expansion.

Even though the ground state energy of the harmonic oscillator does not get effected by this deformation at the first order, we will now demonstrate that
there is a contribution to the energy of the harmonic oscillator from the deformation at second order. The second order correction
to a  general    energy eigen state, from this deformation,   is given by
   \begin{equation}
   \Delta E_{n}^{(2)}=\Sigma_{m \neq n } \frac{|\langle\psi_{m}|\frac{\lambda p}{m}|\psi_{n}\rangle |^2}{E_{n}^{[0]}-E_{m}^{[0]}}.
   \end{equation}
Now for the ground state  $|\psi_{n=0}$, and so we can write the second order correction to the energy of the ground state of the harmonic
oscillator as
   \begin{eqnarray}
   \Delta E_{0}^{(2)}&=&\Sigma_{m\neq 0}\frac{|\langle\psi_{m}|\frac{\lambda p}{m}|\psi_{0}\rangle|^2}{E_{0}^{[0]}-E_{m}^{[0]}} \nonumber \\
&=& \Sigma_{m\neq 0}\frac{|\frac{-\iota\hbar\lambda}{m}\langle\psi_{m}|\frac{d}{dx}|\psi_{0}\rangle|^2}{E_{0}^{(0)}-E_{m}^{(0)}}
\nonumber \\ &=& \Sigma_{m\neq 0}\frac{|\frac{-2\iota\lambda\hbar\alpha}{m} \langle\psi_{m}|x|\psi_{0}\rangle|^{2}}{E_{0}^{(0)}-E_{m}^{(0)}}.
   \end{eqnarray}
Now we can write
   \begin{eqnarray}
  \langle\psi_{m}|x|\psi_{n}\rangle =0,  && m\neq  n\pm 1,
\nonumber \\
  \langle\psi_{m}|x|\psi_{n}\rangle=\sqrt{\frac{n+1}{2\gamma}}, && m=n+1,
\nonumber \\
   \langle\psi_{m}|x|\psi_{n}\rangle=\sqrt{\frac{n}{2\gamma}},&& m=n-1.
   \end{eqnarray}
  where $\gamma= {m\omega}/{\hbar}$. Now the third condition gives an unphysical result, and so we only consider
  $|\psi_{m=1}\rangle$ and $|\psi_{m\neq 1}\rangle$. Now for  $|\psi_{m=1}\rangle$, if
     $E_{0}$ and $E_{1}$ are unperturbed original  ground state and first excited state energies of  the harmonic oscillator given, then we can
     write $ E_{0}={\hbar\omega}/{2}, $ and   $ E_{1}={3\hbar\omega}/{2}, $, so we obtain
     \begin{eqnarray}
         \Delta E_{0}^{(2)}=-\frac{\lambda^{2}}{2m}.
     \end{eqnarray}
However, for  $m\neq 1$, we obtain
 \begin{equation}
 \Delta E_{0}^{(2)}=0.
 \end{equation}
 Thus, there is no second order correction for $|\psi_{m\neq 1}\rangle$, however, the energy of the harmonic oscillator receives a second
 order correction for $|\psi_{m =1}\rangle$. It is interesting to note that various physical systems can be represented by a harmonic oscillator,
 and this includes heavy meson systems like charmonium \cite{char}. The charm mass of this system is $m_c = 1.3 GeV/c^2$. The binding energy of this
 system is approximately equal to the energy gap separating the adjacent levels, which is given by $\hbar \omega \sim 0.3 GeV$. The current
 level of precision measurement is of the order $10^{-5}$ \cite{char2}. Thus, we can use this to set a bound on $\lambda$ as
$   \lambda \leq 10^{-21}.
$ So, the value of $\lambda$ parameterizing this deformation cannot exceed this value, as this bound would violate experimentally known results.

\section{  Landau Level}
In this section, we will analyse the effect of such a deformation on Landau levels.
A charged particles in a magnetic field  can only occupy orbits with discrete energy values due to quantum mechanical effects.
These discrete energy values are called Landau levels. These
  Landau levels are degenerate, and the  number of electrons in a given level is
  directly proportional to the strength of the applied magnetic field. Now we will analyse the effect of
  deforming the momentum operator by $ p\to p(1+\lambda{p}^{-1})$  on Landau levels of a system.
The Hamiltonian for this system  will get corrected by this deformation as
   \begin{eqnarray}
   H &=& \frac{(p-eA)^2}{2m}\to \frac{(p-eA)^2}{2m}+ \frac{\lambda(p-eA)}{m}
   \\ \nonumber &=& H + \lambda H_h,
   \end{eqnarray}
   where $A$ is the vector potential applied to this system.
  We can express the correction term generated from the deformation of this system $H_h$, in terms of the original Hamiltonian $H$
 as
\begin{equation}
 H_h =  \frac{\sqrt{2}\lambda(H)^\frac{1}{2}}{(m)^\frac{1}{2}}.
\end{equation}
 So, the  first order correction to the energy of the  $n$ state can be written as
  \begin{eqnarray}
  \Delta E_{n}{ }&=& \langle\psi_{n}\bigg|\frac{\sqrt{2}\lambda H^\frac{1}{2}}{(m)^\frac{1}{2}}\bigg|\psi_{n}\rangle
\nonumber \\
  &=& \frac{\sqrt{2}\lambda (\hbar\omega)^\frac{1}{2} (n+\frac{1}{2})^\frac{1}{2}}{(m)^\frac{1}{2}}.
   \end{eqnarray}
  Now the  corrections to  energy of this system  is given by
   \begin{equation}
   \frac{\Delta E_{n}{ }}{ E_{n}^{(0)}}=\frac{\sqrt{2}\lambda (\hbar\omega)^\frac{1}{2} (n+\frac{1}{2})^\frac{1}{2}}{(m)^\frac{1}{2} (n+\frac{1}{2})\hbar\omega }             
   \end{equation}
   We can write the corrections to the energy for $n =1$ as
   \begin{eqnarray}
  \frac{\Delta E{_1}{ }}{E_{1}^{(0)}} &=&\frac{\sqrt{2}\lambda}  {\sqrt{\hbar\omega}\sqrt{m}\sqrt{\frac{3}{2}}}
\nonumber \\ &=&\frac{2\lambda}{\sqrt{3\hbar m\omega}}.
  \end{eqnarray}
 Thus, the energy of the Landau levels gets corrected at first order due to the deformation of the momentum operator.
  It may be noted that Landau levels have been determined using the scanning tunneling microscope,
 and for an electron in a magnetic field of $10 T$, we obtain $\omega = 10^3 GHz$, and so the bound on $\lambda$ from the Landau levels is also
of the order $
 \lambda \leq 10^{-22}.
$ This  bound on the
value of $\lambda$ is again obtained using experimental data, and so $\lambda$ greater than this value would violate known experimental results
for Landau levels.

\section{Lamb Shift}
In this section, we will analyse the effect of this deformation on the Lamb shift.
The Lamb shift  occurs due to the interaction between vacuum energy fluctuations
and the hydrogen electron in  different orbitals. This shift can be calculted using quantum theory
of the hydrogen atom, and so we expect that the wave function describing this system will get corrected due
to the deformation of the momentum operator. Thus, we will analyse the effect of this deformation on the   wave function
of such a system.
The   potential energy  of this system can be expressed as $V(r) = -k/r$ and so we can write the deformation of the Hamiltonian
for this system as
\begin{eqnarray}
 H = \frac{p^2}{2m} - \frac{k}{r} \to   \frac{p^2}{2m} - \frac{k}{r} + \frac{\lambda p}{m}.
\end{eqnarray}
To first order,   the wave function of this system  can be expressed as
\begin{eqnarray}
|\psi_{nlm}\rangle_{1}=|\psi_{nlm}\rangle  +
\Sigma_{n^{'}l^{'}m^{'}\neq {nlm}} \frac{e_{n^{'}l^{'}m^{'}|nlm}}{E_{n}^{(0)}-E_{n^{'}}^{(0)}}|\psi_{n^{'}l^{'}m^{'}}\rangle,
\end{eqnarray}
where
\begin{equation}
e_{n^{'}l^{'}m^{'}|nlm}=\langle \psi_{n^{'}l^{'}m^{'}}| \frac{\lambda p}{m} |\psi_{nlm}\rangle.
\end{equation}
Now for the ground state, $n=1, l=0, m=0$, and  the  wave function is given by
\begin{equation}
\psi_{100}=\frac{1}{\sqrt{\pi a_{0}^{3}}} e^{-\frac{r}{a_{0}}}.
\end{equation}
So, for the  first excited state with $l=0$,  we have $m=0$ and $n=2$, and the   wave function can be written as
\begin{equation}
\psi_{200}=\frac{1}{\sqrt{8\pi a_{0}^{3}}}(1-\frac{r}{2a_{0}})e^{-\frac{r}{2a_{0}}}
\end{equation}
The radial  momentum operator can also be expressed  as
\begin{equation}
p=-\frac{\iota\hbar}{r}\frac{d}{dr}(r)=-\frac{\iota\hbar}{r}
\end{equation}
Thus, we obtain
\begin{eqnarray}
e_{200|100} &=& \frac{(-\iota\hbar\lambda)}{m}\int_{0}^{\infty}\int_{0}^{\pi}\int_{0}^{2\pi}
\frac{1}{\sqrt{8\pi a_{0}^{3}}}\Big(1-r/2a_0\Big)\nonumber \\ && \times e^{-r/2a_{0}}\frac{1}{r}\frac{1}{\sqrt{\pi a_{0}^{3}}}e^{-r/a_{0}} r^{2}
\sin\theta dr d\theta d\phi
 \nonumber \\  &=&
 -\frac{\iota\hbar\lambda }{m\sqrt{8\pi a_{0}^{3}}\sqrt{\pi a_{0}^{3}}} \int_{0}^{\infty}r(1-\frac{r}{2 a_{0}})e^{\frac{-3r}{2a_{0}}} dr
 \nonumber \\ && \times \int_{0}^{\pi}\sin\theta d\theta \int_{0}^{2\pi} d\phi
 \nonumber \\  &=&
  \frac{-\iota\hbar\lambda 4\pi}{m a_{0}^{3}  \pi \sqrt{8}} \Big[\int_{0}^{\infty}
  re^{-\frac{3r}{2a_{0}}}dr-\frac{1}{2a_{0}}\int_{0}^{\infty} r^{2}e^{-\frac{3r}{2a_{0}}} dr \Big].
\nonumber \\  &=&
 \frac{-\iota\hbar\lambda\sqrt{32}}{m27a_{0}}
 \end{eqnarray}
 To the  first order, the correction to the    ground state wave function is given by
 \begin{eqnarray}
 \Delta\psi_{100}(r) &=& \psi_{100}^{(1)}-\psi_{100}^{(0)}\\ \nonumber
 &=&\frac{e_{200|100}}{E_{1}^{(0)}-E_{2}^{(0)}} \psi_{200}(r),
 \end{eqnarray}
 where   $E_{n}={-E_{0}}/{n^{2}}$ and  $E_{0}=13.6 eV$.  Thus, we have  $E_{1}=-E_{0}$  and $E_{2}=-{E_{0}}/{4}$
 Thus, the first order correction to the ground state wave function  can be written as
\begin{eqnarray}
\Delta \psi_{100}(r) &=& \frac{-\iota\hbar\lambda\sqrt{32}}{m27a_{0}} \frac{1}{(-E_{0}+\frac{E_{0}}{4})} \psi_{200}(r)
\nonumber \\
&=& \frac{\iota\hbar\lambda\ 8 \sqrt{8}}{m81a_{0}E_{0}} \psi_{200}(r).
\end{eqnarray}
Thus, the wave function for the Lamb shift gets corrected due to the deformation of the momentum operator.
As the Lamb shift depends on the wave function, so a deformation of the wave function, will also deform the Lamb shift.
Hence, the Lamb shift will get corrected at first order due to this deformation of the momentum operator.
The Lamb shift for the $n^{th}$ level is given by
\begin{eqnarray}
\Delta E_n^{(1)} = \frac{4\alpha^2}{3m^2} \left( \ln \frac{1}{\alpha} \right) \left|
\psi_{nlm}(0) \right|^2~.
 \end{eqnarray}
Varying $\psi_{nlm}(0)$, the additional contribution
due to deformation in proportion to its original value \cite{n7}
\begin{eqnarray}
\frac{\Delta E_{n(c)}^{(1)}}{\Delta E_n^{(1)}} =
2 \frac{\Delta|\psi_{nlm}(0)|}
 {\psi_{nlm}(0)}~,
 \end{eqnarray}
 where $\Delta E_{n(c)}^{(1)}$ is the corrected energy due to deformation of the momentum operator.
Thus, for the ground state, the effect of this deformation can be written as
\begin{eqnarray}
\frac{\Delta E_{1(c)}^{(1)}}{\Delta E_1^{(1)}} = 2.7 \lambda \times 10^{23}
 \end{eqnarray}
As the current accuracy of precision in the measurement of Lamb shift is one part in $10^{12}$,
we get the bound on $\lambda \leq 10^{-35}$. Thus, the value of $\lambda$ has to be less than this amount, to be consistent
with present accuracy of measurement of the Lamb shift. It may be noted as the Lamb shift is measured with  more accuracy than
Landau levels, or a system represented by harmonic oscillator, it produces the lowest bound on the value of $\lambda$.
\section{Potential Barrier}
In this section, we will analyse the effect of this deformation on a potential barrier. The potential barrier is important
physically  as it can be used to model different physical systems like the scanning tunneling microscope.
Thus, we will deform the momentum by   $p \to p(1+\frac{\lambda}{p}) $, and analyse its effects on  a  potential barrier.
The deformed  Schroedinger equation for this system can be written as
\begin{equation}
\frac{d^{2}\psi}{dx^{2}}+\Big(\frac{2\iota\lambda}{\hbar}\Big)\frac{d\psi}{dx}-\frac{2m(V_{0}-E)}{\hbar^{2}}\psi=0.
\end{equation}
We will now analyse the solutions to this deformed Schroedinger equation for different regions of this system.

In  the first region, we consider  $V=0$, and the deformed Schroedinger equation in this region can be written as
\begin{equation}
\frac{d^{2}\psi_{1}}{dx^{2}}+\Big(\frac{2\iota\lambda}{\hbar}\Big)\frac{d\psi_{1}}{dx}+\frac{2mE}{\hbar^{2}}\psi_{1}=0.
\end{equation}
The solutions to this deformed Schroedinger equation in this region can be written as
\begin{equation}
\psi_{1}=A e^{\frac{\iota}{\hbar}\Big[\sqrt{\lambda^{2}+2mE}-\lambda\Big]x}+Be^{-\frac{\iota}{\hbar}\Big[\sqrt{\lambda^{2}+2mE}+\lambda\Big]x}.
\end{equation}
Here the  second part   represents a reflected wave from the barrier.
The first part can behave like an incident positive  wave, if it satisfies the  following condition,
\begin{equation}
\bigg|\sqrt{\lambda^{2}+2mE}\bigg|>|\lambda|
\end{equation}
If this condition is not imposed, we will obtain   an unphysical result. Now we can write this solution as
\begin{equation}
\psi_{1}=A e^{\iota k_{1}x}+Be^{-\iota k_{2}x}
\end{equation}
where $k_{1}= (\sqrt{\lambda^{2}+2mE}-\lambda)/{\hbar}$ and $k_{2}= (\sqrt{\lambda^{2}+2mE}+\lambda)/{\hbar}$.
In the second region, we consider  $ V=V_{0}$, and  so the deformed  Schroedinger equation can be written as
\begin{equation}
\frac{d^{2}\psi_{2}}{dx^{2}}+\Big(\frac{2\iota\lambda}{\hbar}\Big)\frac{d\psi_{2}}{dx}-\frac{2m(V_{0}-E)}{\hbar^{2}}\psi_{2}=0.
\end{equation}
The solution to this  deformed  Schroedinger equation can be written as
\begin{equation}
\psi_{2}=C e^{\iota k_{3}x}+De^{-\iota k_{4}x}.
\end{equation}
where $k_{3}= (\sqrt{\lambda^{2}-2m(V_{0}-E)}-\lambda)/{\hbar}$ and $k_{4}= (\sqrt{\lambda^{2}-2m(V_{0}-E)}+\lambda)/{\hbar}$.
The only difference between the solution in the third region and solution in the first region is that there is no
 reflected wave in the third region. So,  the solution to the deformed Schroedinger equation in the third region can be written as
\begin{equation}
\psi_{3}=Ee^{\iota k_{1}x}
\end{equation}
where $k_{1}= \ (\sqrt{\lambda^{2}+2mE}-\lambda)/ \hbar$.

The most important thing for such systems is the transmission coefficient $T$, and we want to analyse the effect of this deformation
of the momentum on the transmission coefficient of this system. Thus, we will now analyse the effect of this deformation on the
 incident  current density $J_{I}$ and the transmitted current density  $J_{T}$,
\begin{eqnarray}
J_{I}&=& \frac{\hbar k_{1}}{m}|A|^{2}.
\nonumber \\
J_{T}&=& \frac{\hbar k_{1}}{m}|E|^{2}
\end{eqnarray}
Now  to the first order in $\lambda$,   the value of constants $k_{1},k_{2},k_{3},k_{4}$
can be written as
\begin{eqnarray}
k_{1}&=&\frac{1}{\hbar}\Big(\sqrt{2mE}-\lambda\Big),
\nonumber \\
k_{2}&=&\frac{1}{\hbar}\Big(\sqrt{2mE}+\lambda\Big),
\nonumber \\
k_{3}&=&\frac{1}{\hbar}\Big(\sqrt{2m(V_{0}-E)}-\lambda\Big),
\nonumber \\
k_{4}&=&\frac{1}{\hbar}\Big(\sqrt{2m(V_{0}-E)}+\lambda\Big).
\end{eqnarray}
Thus, using the standard analysis for the barrier potential, the effect of the deformation on the potential barrier, will be
given by
\begin{eqnarray}
\frac{E}{A}&=&\Big[    e^{-\iota(k_{1}+k_{4})a}       \Big(k_{3}k_{1}+k_{1}k_{4}\Big)\Big]\nonumber \\&& \times
\Big[   (k_{3}-k_{1})\bigg[k_{2}(k_{1}+k_{4})     \Big(1-e^{-\iota(k_{3}+k_{4})a}\Big)  \nonumber \\&&
-k_{4}(k_{1}+k_{2})    \bigg]+            k_{3}(k_{1}+k_{4})e^{-\iota(k_{3}+k_{4})a}\Big]^{-1}.
\end{eqnarray}
It may be noted that if $T_0$ is the original transmission coefficient  for the potential barrier, and $T$ is the transmission coefficient
for the potential barrier obtained by deforming the coordinate representation of the momentum operator, then we can write
\begin{equation}
T=\frac{J_{T}}{J_{I}}=\bigg|\frac{E}{A}\bigg|^{2}.
\end{equation}
Furthermore, if $I_0$ is the original tunneling current, and $I$ is the tunneling current for the deformed system, then we can write \cite{n7}
\begin{equation}
\frac{I}{I_0}= \frac{T}{T_0}= \frac{1}{T_0} \bigg|\frac{E}{A}\bigg|^{2}.
\end{equation}
So, we expect an excess tunneling current generated from the deformation of this system,
\begin{equation}
 \frac{{I} - I_0}{I_0} =\left[ \frac{1}{T_0} \bigg|\frac{E}{A}\bigg|^{2} -1\right].
\end{equation}
This excess tunneling current can be detected experimentally by using precise experiments, if such a deformation
of this system exists. Thus, this excess tunneling current can be used to test the effects of this deformation proposed
in this paper.
\section{Conclusion}
In this paper, we proposed the   most general form of the generalized uncertainty principle.
It is known that the generalized uncertainty principle deforms the coordinate representation of the momentum operator. Thus, we
construct the most general form of such a deformation of the momentum operator.  Such a general deformation of the momentum operator
contains both fractional derivative terms, and nonlocal terms which can be expressed as kernels of some integral operator.
We also analyse a specific limit of this most general form of the deformation  of the momentum operator, for one dimensional systems. In this limit,
the momentum operator contains nonlocal terms, however, the quantum mechanical Hamiltonian for all one dimensional systems is local.

We analyse the effect of the specific  deformation on a harmonic oscillator and observe that its there is no correction to the  energy
of the harmonic oscillator at first order. However, the energy of the harmonic oscillator does get corrected at second order.
We   analyse the corrections to the energy of Landau levels, and observe that Landau levels gets correct at first order due to this deformation
of the momentum operator. The wave function describing the Lamb shift also gets corrected at first order of the perturbation theory.
We also observe that the transmission coefficient of a barrier potential gets modified due to this deformation of the momentum operator.
Finally, we calculate the effect of this deformation on the particle in a box. We observe that  no particle can exist in a box, if the
length of the box is not quantized. We used this to argue that the space is quantized in  terms of discrete units.
 It is interesting to note that unlike the previous linear deformation, in this deformation the discretization of length depends on the energy
 of a system. Such a dependence of the structure of spacetime on the energy used to probe it is the basis
 of gravity's rainbow \cite{rainbow0}-\cite{rg}. It may be noted that gravity's rainbow has been motivated from string theory \cite{gr},
 and string theory has also been also used as a motivation for generalized uncertainty principle \cite{z2}-\cite{2z}, so it is expected that
 some form of generalized uncertainty principle can produce results  similar to gravity's rainbow.
  Furthermore, gravity's rainbow
 has been used to explain the hard spectra from gamma-ray
burster's~\cite{3q}.
It would be interesting to investigate the relation between this formalism and gravity's rainbow further.
As the deformation studied in the paper can produce conclusions similar to gravity's rainbow, it might be possible
that the deformation used in the paper might also help explain the hard spectra from gamma-ray
burster's.

The deformation of the coordinate representation of the momentum operator will deform all quantum mechanical systems, including
the first quantized field theories. It may be noted that the field theories motivated by the generalized uncertainty principle
have been studied \cite{mir}-\cite{mir1}. It was observed that the first quantized equations of motion for such field theories
gets deformed due to the deformation of the Heisenberg algebra. It would be interesting to perform a similar analysis for
field theories deformed using the deformation proposed in this paper. It is expected that such a deformation will give rise to
non-local terms. Furthermore, it would be interesting to analyse the gauge symmetry corresponding to such non-local gauge theories.
It is known that the non-local gauge theories are usually invariant under a non-local gauge transformation. Thus, we expect
that the gauge theories obtained from such a deformation of field theories would be invariant under non-local gauge transformations.
It would be interesting to analyse the effect of non-locality on different processes and
amplitudes in these non-local   theories. These non-local gauge theories can be used to
analyse the effects of    non-locality on the one-loop amplitudes
and renormalization group flow. Finally, we can also analyse some formal aspects of such theories. So, we can analyse the BRST
quantization of these deformed non-local gauge theories. We expect that as these gauge theories would be invariant under
a non-local gauge transformation, the BRST symmetry for these gauge theories would also contain non-local terms. It would be
interesting to analyse the effect of such non-locality on the BRST symmetry of this theory.

It is also possible to incorporate the generalized uncertainty principle in Lifshitz field theories  \cite{mir1}.
As we have proposed a new deformation of the momentum operator, it would be interesting to incorporate such a deformation
of the momentum operator in field theories based on Lifshitz scaling. It is expected that the deformation parameter would break
the Lifshitz scaling. However, such a parameter can be promoted to a background field, and this field can be made to transform in the
appropriate way to preserve the Lifshitz scaling. It has been observed that
the van der Waals and Casimir interaction between graphene
and a material plate can be analysed using Lifshitz scaling  \cite{a5}.  In fact, the van der Waals and Casimir
interaction between a single-wall carbon nanotube and a
plate can also be analysed    using Lifshitz scaling  \cite{a5}.   It would be interesting to analyse the deformation of this system
by the generalized uncertainty principle. It may be noted that   Lamb shift \cite{lb}-\cite{bl} and Landau levels \cite{la}-\cite{al} have
 been recently studied in graphene, and so it would be interesting to analyse the effects of generalized uncertainty principle on  Landau levels and
 Lamb shift in graphene.

 \section*{Acknowledgments}
 We would like to thank Saurya Das for useful discussions and suggestions.
 
\end{document}